\documentclass[showpacs,showkeys,aps,prc,amsmath,amssymb]{revtex4}

\usepackage{dcolumn}
\usepackage{bm}
\usepackage{epsfig}
\newcommand{\pbar}{\overline{p}}
\newcommand{\mtilde}{\widetilde{m}}

\begin{document}

\title{Probing Matter Radii of Neutron-Rich Nuclei by Antiproton Scattering}
\author{Horst Lenske$^1$ and Paul Kienle$^2$ }%
\affiliation{%
  $^1$Institut f\"ur Theoretische Physik, Universit\"at Gie\ss en,
  Heinrich-Buff-Ring 16, D-35392 Gie\ss en, Germany\\
  $^2$Fakult\"at f\"ur Physik, Technische Universit\"at M\"unchen, James Franck Str. 1, D-85748
  Garching, Germany, and Stefan Meyer Institut f\"ur subatomare Physik, Boltzmanngasse 3, A-1090 Wien,
  Austria
  }%

\date{\today}

\begin{abstract}
We propose to use antiprotons to investigate the sizes of stable and
neutron-rich exotic nuclei by measurements of the $\pbar A$
absorption cross section along isotopic chains in inverse
kinematics. The expected effects are studied theoretically in a
microscopic model. The $\pbar U$ optical potentials are obtained by
folding free space $\pbar N$ scattering amplitudes with HFB ground
state densities and solving the scattering equations by direct
integration. The mass dependence of absorption cross sections is
found to follow closely the nuclear root-mean-square radii. The
total absorption cross section is shown to be a superposition of
cross sections describing partial absorption on neutrons and
protons, respectively. Thus measuring the differential cross
sections for absorption on neutrons and protons will give
information on their respective distributions. In neutron-rich
nuclei the outer neutron layer shields the absorption on the protons
giving access to investigations of antiproton-neutron interactions
in matter.
\end{abstract}

\maketitle

\section{Introduction}

To understand the evolution of nuclear sizes and shapes from the
bottom to the edges of the valley of $\beta-$ stability is one of
the central questions of modern nuclear structure physics.
Experimentally, a variety of efforts is undertaken or in planning to
investigate properties of neutron-rich exotic nuclei ranging from
high-energy breakup reactions \cite{FRS:03} and elastic proton
scattering \cite{Egelhoff:03} to charge-exchange reactions
\cite{Krasznahorkay:99} and low-energy transfer reactions
\cite{LS:97,Jeppsen:04}. Here, we propose to use antiprotons to
probe nuclear sizes of stable and unstable nuclei in a systematic
way.

Investigations of radii and density distributions of stable nuclei
by antiprotons are by itself a long discussed and applied method,
e.g. at BNL \cite{Bugg:73,Ashford:84} or LEAR
\cite{Janouin:86,Bruge:86,Lubinski:94,Wycech:96}. Hitherto, the
experiments have been performed with secondary antiproton beams on a
variety of stable nuclei in fixed target geometry and standard
kinematics, actually using stopped antiprotons. Obviously, another
approach must be applied if antiprotons should be used for reactions
on short-lived isotopes. These nuclei by themselves are available
also only as secondary beams, produced either by fragmentation or
isotope separation on line. A solution dissolving these conflicting
conditions is to perform the measurements in colliding beam
geometry. Such a setup was recently proposed in \cite{PK:04}. It
will become feasible with the meanwhile approved FAIR facility a GSI
\cite{FAIR}. The FAIR plans include as central components $\pbar$
production and accumulation facilities. For nuclear structure
research the New Experimental Storage Ring (NESR) for short-lived
nuclides and an intersecting electron accelerator for colliding beam
experiments will be available \cite{Koop:02}. As discussed in the
Antiproton-Ion-Collider (AIC) proposal \cite{AIC:05}, with moderate
modifications on an acceptable level the NESR/$e^-$ setup can be
converted into a $\pbar A$ collider facility. Details of the
experimental setup and procedures for the measurement of antiproton
absorption cross sections using Schottky noise frequency
spectroscopy of the coasting reaction products in the NESR are found
in the NUSTAR letter of intent and proposal for FAIR
\cite{LoI,PK:04}.

The physics of $\pbar A$ interactions is by itself an interesting
topic for nuclear and hadron physics. It was brought to the
attention of nuclear physics originally by Duerr and Teller
\cite{Duerr:56}. The usefulness of $\pbar+A$ scattering for studies
of the neutron skin in stable medium and heavy nuclei was realized
some years later in an experiment at BNL \cite{Bugg:73}. In the
aftermath, a variety of experiments have been performed, e.g.
\cite{Janouin:86,Bruge:86,Lubinski:94,Wycech:96}. As for the
elementary antiproton nucleon ($\pbar N$) vertex the $\pbar A$
reactions are dominated by processes in which the incoming
antiproton annihilates on a target nucleon into a variety of
particles producing typically and preferentially a shower of pions
as the final result. Here, we utilize the strong $\pbar A$
absorption for nuclear structure investigations without paying
particular attention to the hadron physics aspects, some of which,
however, can -- and in another context will -- be studied with the
facility under discussion, e.g. with the PANDA detector, also being
part of the FAIR proposal.

For our purpose we are satisfied to know that the elementary
annihilation processes result finally in a strong suppression of the
incoming flux. In the elementary elastic scattering amplitude
$f_{\pbar N}$ the annihilation channels appear as a strong imaginary
part while only a small, almost vanishing, real part is observed.
Correspondingly, in potential models for inclusive $\pbar A$ elastic
scattering the annihilation channels are typically accounted for
globally by a strongly absorptive optical potential, as e.g. in
\cite{Suzuki:85,Heiselberg:86,Adachi:87,Zhang:96}. In fact, the
total cross section is about twice the elastic cross section
\cite{Green:84} which underlines the importance of the annihilation
channels.

The new aspect of the present paper is to explore theoretically the
perspectives of antiproton scattering on stable and rare nuclei and
the use of such reactions for nuclear structure investigations. In
particular, we consider elastic $\pbar A$ scattering and study to
what degree the total reaction or absorption cross section provides
information on the size and shape of the target nucleus. The
theoretical approach is described in sect. \ref{sec:Theory}. As a
typical and interesting case absorption cross sections for the $Ni$
isotopic chain are discussed in sect.\ref{sec:Results}. The paper
closes with a summary in sect.\ref{sec:sum}.

\section{Microscopic Description of Antiproton Absorption on Nuclei}\label{sec:Theory}

\subsection{Optical Model Approach for $\pbar A$ Elastic Scattering}

The basic ingredients for a microscopic description of $\pbar A$
scattering are obviously the elementary antiproton-nucleon
interactions. Including a spin-independent amplitude $A(\sqrt{s},t)$
and a spin-orbit part $C(\sqrt{s},t)$ the free space antiproton
scattering amplitude is taken to be
\begin{equation}
f_{\pbar N}(\sqrt{s},q) = A(\sqrt{s},t) +
C(\sqrt{s},t)\frac{1}{k}\mathbf{\sigma}_{\pbar} \cdot (\mathbf{q}
\times \mathbf{k}) \quad ,
\end{equation}
where $k$ is the momentum of the incoming antiproton, s the
Mandelstam center of mass energy and $q^2=-t$ is the momentum
transfer. For kinetic energies T$_{lab} \lesssim 1500$~MeV the data
clearly show that the $\pbar N$ interactions are dominated by
annihilation processes. By means of the optical theorem an effective
T-matrix is defined
\begin{equation}\label{eq:tparam}
t_{\overline{p}N}(T_{Lab},q^2) = \frac{2\pi\hbar}{M}
\frac{ik}{4\pi}\sigma_{\overline{p}N}(T_{Lab})\left(1-i\epsilon\right)F_{\overline{p}N}(q^2)
\end{equation}
where $T_{Lab}=E_{Lab}(k)-M$ is the kinetic energy of the incoming
antiproton with momentum $k$ and the total $\overline{p}N$ cross
section $\sigma_{\overline{p}N}(T_{Lab})$. The ratio of the real and
the imaginary part is denoted by $\epsilon$ which lies between
$\frac{1}{4} \cdots \frac{1}{3}$ in the energy range considered
here. As in \cite{Tan:89} we use a t-channel form factor of gaussian
shape, $F(q^2)\equiv e^{-\beta^2q^2}$. A closer inspection shows
that by eq.\ref{eq:tparam} $t_{\pbar N}$ is assumed to be separable
into an on-shell strength factor and an off-shell form factor.

The coordinate space $\pbar A$ optical potential is then obtained in
the {\em impulse} approximation and after a Fourier transformation,
\begin{equation}\label{eq:fold}
U_{opt}( \mathbf{r}) = \sum_{N=p,n}{ \int{\frac{d^3q}{2\pi^3}
\rho_N(q)t_{\overline{p}N}(T_{Lab},q^2)e^{i\mathbf{q\cdot r}}}}
\quad ,
\end{equation}
a strongly absorptive complex optical potential $U_{opt} = V + i W$
with real and imaginary parts $V$ and $W$, respectively, is
obtained. Obviously, we have $U_{opt}=U^{(p)}_{opt}+U^{(n)}_{opt}$.

In practice, we use the momentum space amplitudes, including also a
complex spin-orbit part, derived in \cite{Tan:89} according to the
Paris model \cite{Cote:82}. As in the elementary $\pbar N$
interaction the potentials are dominated by strong imaginary parts.
We also include the (long range) Coulomb potential $U_c$ which is
calculated microscopically by folding the elementary $\pbar N$
Coulomb amplitude with the HFB charge density $\rho_c$. The
elementary proton and neutron electric form factors
$G^{(n,p)}_E(q^2)$ are taken into account. The $\pbar A$ spin-orbit
optical potentials $U_{so}$ are calculated with the HFB spin
densities. Although the resulting $U_{so}$ are rather weak and, in
particular, do not contribute significantly to the absorption they
are taken into account in the calculations for completeness.

We use a non-relativistic approach and describe $\pbar A$ scattering
by the optical model Schroedinger equation
\begin{equation}\label{eq:WaveEq}
\left( -\frac{\hbar^2}{2\mtilde}\overrightarrow{\nabla}^2 +
U_{opt}-T_{cm}\right)\Psi^{(+)}_{\alpha}(\mathbf{k},\mathbf{r})=0
\quad .
\end{equation}
Here, $\Psi^{(+)}_{\alpha}(\mathbf{k},\mathbf{r})$ is the optical
model wave function for the incoming proton in spin state
$m_\alpha=\pm \frac{1}{2}$ and with asymptotically outgoing
spherical waves. The antiproton-nucleus center-of-mass system with
reduced mass $\mtilde$ is used.

From the solutions of eq.\ref{eq:WaveEq} we determine the partial
wave S-matrix elements $S_{\ell j}$ \cite{Joachain} where for an
absorptive system $|S_{\ell j}|<1$. The continuity equation for the
optical model wave function $\Psi^{(+)}$ provides us with the exact
relation
\begin{equation}\label{eq:xsabs}
\sigma_{abs} = {\frac{\pi}{k^2}\sum_{\ell
j}\frac{2j+1}{2s+1}(1-|S_{\ell
j}|^2)}=-\frac{2\mtilde}{k\hbar^2}\frac{1}{2s+1}tr_s\left(\int{d^3r\Psi^{(+)\dag}(\mathbf{k},\mathbf{r})
\Im{(U_{opt})}(\mathbf{r})} \Psi^{(+)}(\mathbf{k},\mathbf{r})\right)
\quad ,
\end{equation}
with averaging over the $\pbar$ spin projections. The second
equality provides us with a particular useful relation: From
eq.\ref{eq:fold} we know $\Im{(U_{opt})}=W_p+W_n$ which immediately
implies additivity for $\sigma_{abs}=\sigma^{(p)}_{abs} +
\sigma^{(n)}_{abs}$  and the partial cross section for annihilation
on $q=p,n$, respectively, are given by
\begin{equation}\label{eq:xsabspn}
\sigma^{(q)}_{abs}=-\frac{2\mtilde}{k\hbar^2}\frac{1}{2s+1}tr_s\left(\int{d^3r|\Psi^{(+)\dag}(\mathbf{k},\mathbf{r})|^2
W_q(\mathbf{r}) }\right) \quad .
\end{equation}

\begin{figure}[htb]
\begin{center}
\begin{minipage}[b]{14 cm}
\begin{center}
\epsfig{file=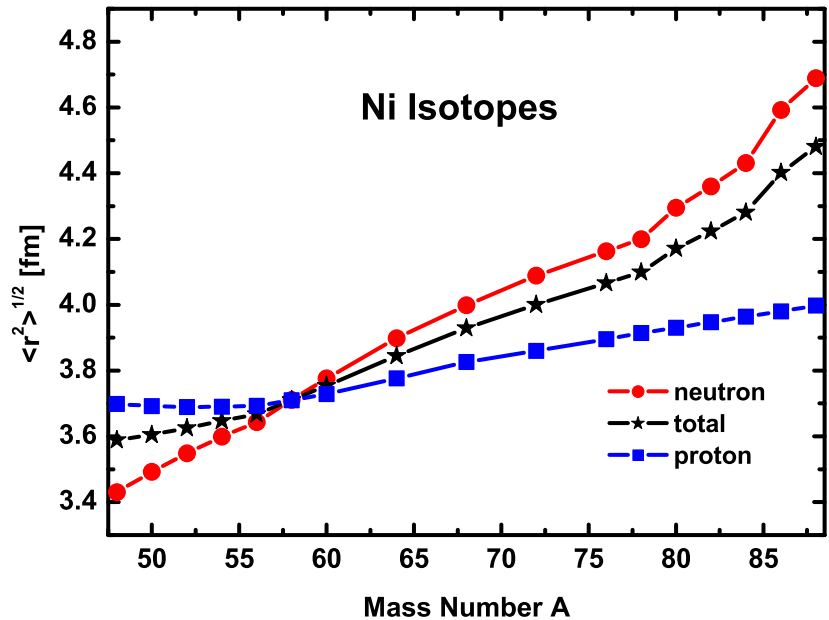,width=13cm}
\end{center}
\end{minipage}
\begin{minipage}[b]{14 cm}
\caption{Root-mean-square (rms) radii for $^{48-88}$Ni-isotopes. HFB
results for protons (squares), neutrons (circles) and the total
(isoscalar) density (stars) are shown. The mass range corresponds to
the isotopes found in the HFB calculations to be particle-stable.
The lines are to guide the eye. Note the change from proton skins at
$A\lesssim 58$ to neutron skins at larger mass numbers.
}\label{fig:rms}
\end{minipage}
\end{center}
\end{figure}

For our aim eq.\ref{eq:xsabspn} is of particular interest because
the leading order proton and neutron contributions to $\sigma_{abs}$
are being displayed explicitly. Hence, we have succeeded to
represent the annihilation on the target protons and neutrons in a
formally separable form allowing to study theoretically the
corresponding partial absorption cross sections separately. However,
we note that in eq.\ref{eq:xsabspn} the distorted waves are
determined by $U_{opt}$ in a non-perturbative manner and, as such,
include the whole multitude of effects from $W_{p,n}$.

The ground state densities entering into the optical potentials are
taken from non-relativistic HFB calculations using a G-Matrix
interaction as described in \cite{D3Y:98}. The known deficiencies of
non-relativistic Brueckner-calculations are overcome by introducing
additional density dependent contributions \cite{D3Y:98,Le:02} and
adjusting the parameters to the variational results for infinite
nuclear matter of the Urbana group \cite{Akmal:98}. An effective
density dependent zero-range pairing interaction is derived from the
SE (S=0,T=1) in-medium NN-interaction \cite{Le:02}. The system of
coupled state dependent HFB gap equations is solved
self-consistently. HFB results for rms-radii of the proton and
neutron ground state densities in the $Ni$ isotopes are displayed in
Fig.\ref{fig:rms}. We also find a very satisfactory overall
agreement of the theoretical and the measured binding energies on
the level of 5\%, see e.g. \cite{Le:02,Nadia:04}

\subsection{Absorption Cross Sections for $\pbar$ Annihilation on the
Ni-Isotopes}\label{sec:Results}

While for most $\pbar A$ investigations the strong absorption is an
unwanted and disturbing  effect, we take advantage of this property:
Because of the strong absorption the product of scattering waves in
eq.\ref{eq:xsabspn} gains strength mainly in the nuclear surface
thus favoring investigations of nuclear skin configurations. As seen
from eq.\ref{eq:xsabspn} this behavior, common for both the proton
and neutron partial cross sections, is modulated by the specific
shape of $W_{p,n}$ leading to a pronounced weighting of the nuclear
surface region. Thus, measurements of the $\pbar A$ absorption cross
section will provide us with data proving the existence of nuclear
skins and recording their properties like their composition and mass
dependence.

\begin{figure}[htb]
\begin{center}
\begin{minipage}[htb]{14 cm}
\begin{center}
\epsfig{file=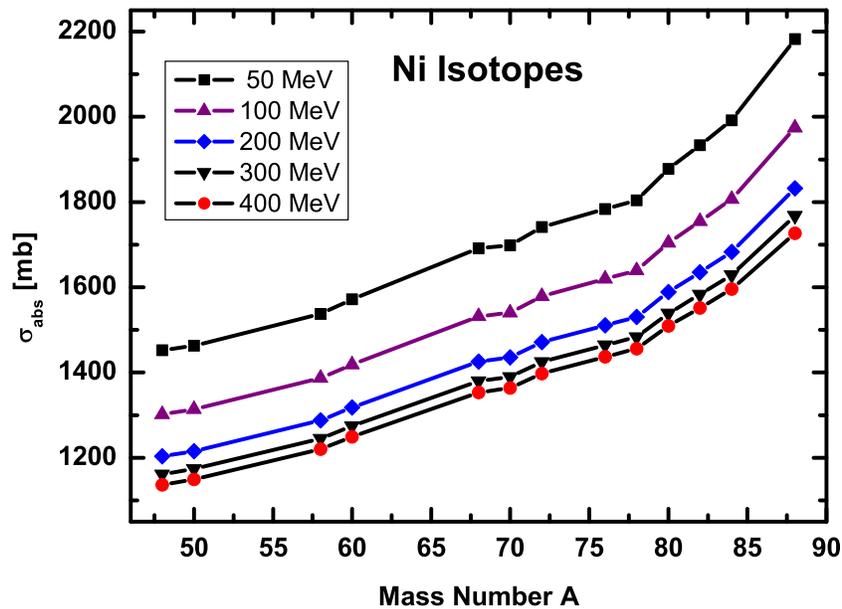,width=13cm}
\end{center}
\end{minipage}
\begin{minipage}[htb]{14 cm}
\caption{Absorption cross sections for antiproton annihilation on
the $^{48-88}$Ni-isotopes at various antiproton incident energies.
The lines are to guide the eye.}\label{fig:sigABS}
\end{minipage}
\end{center}
\end{figure}

The total absorption cross sections for Ni-Isotopes in the mass
range 48$\le$A$\le$88 are shown in Fig.\ref{fig:sigABS}. Results for
antiproton incident energies from T$_{Lab}=$50~MeV to
T$_{Lab}=$400~MeV are displayed. With increasing energy the
magnitude of $\sigma_{abs}$ decreases at a rate slowing down when
approaching the highest shown energy, $T_{lab}=400$~MeV. A
remarkable feature common to all energies is the steady increase
with mass number, reflecting the growth of the neutron skin with the
neutron excess.

In Fig.3 the absorption cross sections at T$_{Lab}$=200~MeV and
T$_{Lab}$=400~MeV are compared to the rms-radii of the nuclear
ground state number density distributions.  At both energies the
nuclear rms-radii were normalized arbitrarily to the cross sections
at $^{58}$Ni expressed by the scaling law $\sigma_{abs}\simeq
\alpha_0 \langle r^2\rangle$. From Fig.3 we note an impressive
agreement for $\sigma_{abs}$ and the rms radii, both increasing with
almost the same slope as functions of the target mass. This behavior
is found at all energies considered here, thus indicating an almost
universal scaling law . The normalization constants $\alpha_0$ are
slowly varying with the incident energy, decreasing by about 5\%
from $T_{Lab}=200$~MeV to $T_{Lab}=400$~MeV. Their origin and
physical content will be discussed below. We conclude that
antiproton annihilation is a well suited probe for investigating
nuclear sizes and shapes.

\begin{figure}[htb]
\begin{center}
\epsfig{file=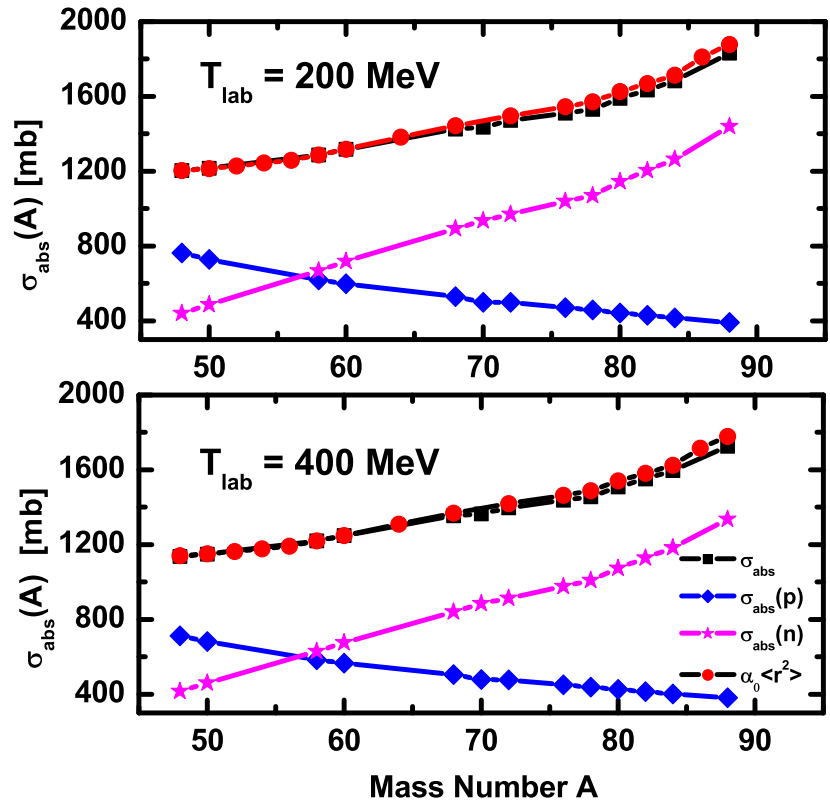,width=13cm} \caption{Absorption cross sections
(filled squares) for antiproton annihilation on Ni-isotopes at
$T_{lab}=200$~MeV and $T_{lab}=400$~MeV, respectively, are compared
to the rms-radii of the nuclear matter densities (filled circles).
In both cases, the rms-radii are arbitrarily normalized by a factor
$\alpha_0$ to $\sigma_{abs}$ for $^{58}$Ni. Also shown are the
partial cross sections for absorption on the target neutrons (stars)
and protons (diamonds).}
\end{center}\label{fig:sigRMS}
\end{figure}

It is instructive to inspect $\pbar A$ scattering along a trajectory
through the target nucleus by means of the eikonal approximation
\cite{Joachain}. The partial absorption cross sections for $q=p,n$
at impact parameter $b$ are given by
\begin{equation}\label{eq:eikonal}
\sigma^{(q)}_{abs}(b) =\frac{2\pi}{k}\frac{\mtilde}{
\hbar^2}\int^{+\infty}_{-\infty}{dz\Pi_q(z,b)} \quad ,
\end{equation}
with the eikonal absorption kernel
\begin{equation}\label{eq:EKernel}
\Pi_q(z,b)=\frac{1}{2s+1}tr_s\{\Pi^{(s)}(z,b)\}=-e^{-2\chi_q(z,b)}
W_q(z,b) \quad ,
\end{equation}
including a trace over spin projections. The eikonal integral
\begin{equation}\label{eq:profile}
\chi_q(z,b)= -\frac{2\mtilde}{k \hbar^2}\int^{z}_{-\infty}{d\xi
W_q(\xi,b)}
\end{equation}
describes the depletion of the incoming antiproton flux along the
trajectory through the target nucleus. These relations show that
$\Pi(z,b)$ is a measure for the degree of absorption at a given
point in the $(z,b)$ scattering plane. Overall, for impact
parameters $b$ less than the target radius $R_A$ the incoming flux
is strongly absorbed, giving the target the characteristics of a
black disk.

\begin{figure}[htb]
\begin{center}
\epsfig{file=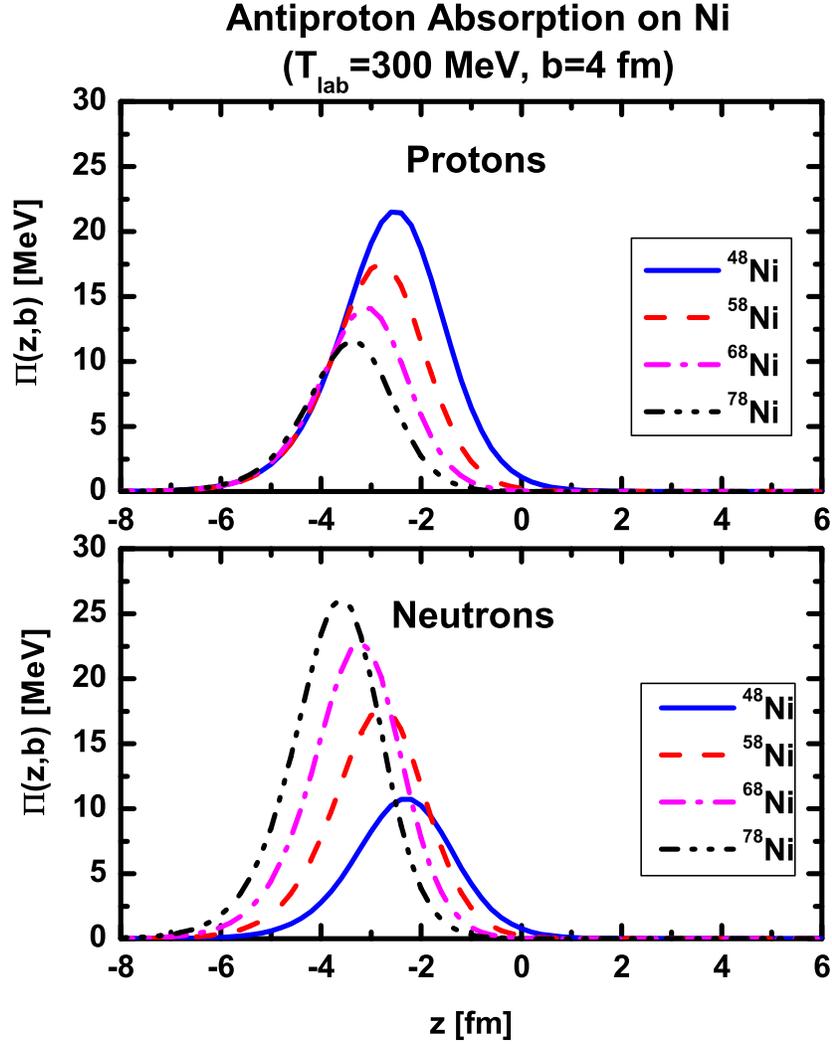,width=13cm} \caption{Absorption of the
incoming antiproton flux along a trajectory with impact parameter
$b=4$~fm. The eikonal absorption kernel,
eq.\protect\ref{eq:EKernel}, is displayed for annihilation of
antiprotons with an incident energy of $T_{lab}=300$~MeV on
$^{48,58,68,78}$Ni. The $\pbar$ beam is incoming from the left, i.e.
from $z=-\infty$. Note the reversed behavior of the $\pbar$
absorption on protons and neutrons with increasing neutron excess.}
\end{center}\label{fig:Kernel}
\end{figure}

In Fig.4, results for $\Pi(z,b)$ are displayed at $T_{Lab}=300$~MeV
for a variety of $Ni$ nuclei and an impact parameter $b=4$~fm which
roughly corresponds to grazing collisions over the mass region
considered in Fig.4. We find an interesting mass dependence: In the
neutron-deficient $^{48}$Ni the absorption is takes place dominantly
in the {\bf proton skin} and with increasing neutron number the
process changes gradually to absorption on the neutrons with a
break-even point around $^{58}$Ni. In the extremely neutron-rich
isotopes beyond $^{78}$Ni the antiprotons are absorbed
preferentially in the neutron skin, i.e. in the outer layers of the
nuclear density distribution. As a consequence, in the neutron-rich
isotopes the more tightly bound and spatially more confined protons
are effectively screened from the incoming $\pbar$ flux by the
neutron skin. This implies that the antiprotons are preferentially
annihilated in a region of almost pure neutron matter. The screening
effect also explains the decrease of the partial proton and the
increase of the partial neutron optical model cross
sections,respectively, seen in Fig.3. In fact, the eikonal approach
leads to a very satisfying agreement with the exact optical model
results even on the quantitative level. Above $T_{Lab}\simeq
100$~MeV both descriptions agree by better than 10\% and, as to be
expected, the deviations diminish with increasing incident energy.

Assuming a gaussian form factor for $W(r)$ the eikonal cross
section, eq.\ref{eq:eikonal}, can be calculated in closed form,
resulting in $\sigma_{abs}(A,T_{cm})=f(A,T_{cm})\pi \langle
r^2\rangle_A$. The eikonal scaling function
\begin{equation}\label{eq:scaling}
f(A,T_{cm}) = \gamma+\log{(\xi(A,T_{cm}))}+Ei(\xi(A,T_{cm}))
\end{equation}
describes the deviation from the black disk limit
$\sigma_{abs}\to\pi \langle r^2\rangle_A$. $\gamma=0.5772\cdots$
denotes Euler's constant, $Ei(x)$ is the exponential integral
\cite{AS:65} and
\begin{equation}\label{eq:arg}
\xi(A,T_{cm})=\frac{\sqrt{\pi}\ell_g \Im{(t_{\pbar
N}(T_{cm}))}\rho_0}{T_{cm}} \quad .
\end{equation}
is given in terms of physical quantities, including the $\pbar N$
T-matrix, the central nuclear density $\rho_0$ and $\ell_g = kR_A$
corresponding to the $\pbar A$ grazing angular momentum. For the
physically relevant values of $\xi$ the $Ei(\xi)$ term can safely be
neglected. These relations enable us to interpret in physical terms
the scaling factor introduced above. By comparison we find $\alpha_0
\simeq a_0=\frac{3\pi }{2}f(A=58,T_{cm})$. Indeed, approximating the
microscopic $W(r)$ by a Gaussian of the same volume integral and
rms-radius the analytic eikonal results agree with the full optical
model $\alpha_0$ convincingly well, e.g. $\alpha_0/a_0 \simeq
0.92\cdots 0.94$ for $T_{Lab}=100\cdots 400$~MeV and unity is
approached a higher energies. The functional structure of the
eikonal scaling factor, eq.\ref{eq:scaling}, also explains the
apparent convergence of the absorption cross sections to limiting
asymptotic values as indicated in Fig.\ref{fig:sigABS} for the
larger energies. Thus, $f(A,T_{cm})$ and correspondingly $\alpha_0$
contain valuable information on the dynamics of the $\pbar A$
system.

\section{Conclusions and Outlook}\label{sec:sum}

Antiproton-nucleus scattering was investigated theoretically as a
tool for measuring neutron skins in exotic nuclei. The $\pbar A$
absorption cross sections were calculated with microscopic optical
potentials derived from the elementary free space $\pbar N$
scattering amplitudes and HFB density distributions. The results
point to promising perspectives in at least two directions, namely
\begin{itemize}
\item showing that antiprotons are a perfect probe for
investigations of neutron (and proton) skins;
\item indicating a new method for investigating
antiproton-neutron interactions.
\end{itemize}
The latter point is of particular interest because the data base on
$\pbar n$ interaction is very scarce. It also adds a new aspect to
the experiment proposed in \cite{PK:04,AIC:05}: Scanning over an
isotopic chain will allow to record the gradual change from $\pbar
p$ to $\pbar n$ annihilation, hence providing a new access to the
$pbar$-nucleon interactions.

Our results show that simple absorption cross section measurements
will already provide valuable new information on the densities of
the target nuclei, most likely of a higher accuracy than achievable
by other methods. More involved measurements seem to be feasible
from which more detailed information on the configurations of the
target nuclei can be obtained.

A next step will be to consider more exclusive experiments.
Differential measurements by gating on the $A-1$ residual nuclei
which will provide information on the annihilation of the incoming
$\pbar$ on a single target neutron or proton, respectively, are
being explored \cite{AIC:05}. Parallel to the nuclear structure
studies the experimental program should also envision investigations
of elastic $\pbar A$ scattering. A consistent and much extended set
of elastic angular distributions is highly desired by theory for a
better understanding of antiproton interactions in nuclear
matter.\vspace{2mm}

\noindent Supported in part by DFG, contract Le 439/05, GSI and
BMBF.

\end{document}